\documentstyle[prb,aps,epsf]{revtex}
\def\-{\hphantom{$-$}}
\def\0{\hphantom{0}}
\def\ket#1{|#1\rangle}
\def\k{{\bf k}}
\def\R{{\bf R}}
\def\S{{\bf S}}
\def\i{{\rm i}}
\begin{document}
\title{Direct exchange in the edge-sharing spin-1/2 chain compound MgVO$_3$} 
\author{I. Chaplygin, R. Hayn }
\address{Institute for Theoretical Physics, TU Dresden, D-01062
Dresden, Germany}
\author{K. Koepernik}
\address{Max Planck Institute CPFS, D-01187 Dresden, Germany}
\date{\today}

\twocolumn[\hsize\textwidth\columnwidth\hsize\csname@twocolumnfalse\endcsname
\maketitle
\begin{abstract}
Bandstructure calculations with different spin arrangement for the 
spin-chain compound MgVO$_3$ have been performed, and paramagnetic as well as 
magnetic solutions with ferro- and antiferromagnetically ordered chains are 
found, the magnetic solutions being by 0.22 eV per formula unit lower than the
paramagnetic one. The orbital analysis of the narrow band crossing the Fermi
level in the paramagnetic solution reveals that the band has almost pure 
vanadium $3d$ character, the lobes of the relevant $d$-orbitals at the 
neighboring in-chain sites being directed towards each other, which suggests 
direct exchange. The tight-binding analysis of the band confirms the strong 
exchange transfer between neighboring in-chain V-ions. Besides, some 
additional superexchange transfer terms are found, which give rise both to 
in-plane coupling between the chains and to frustration, the dominant 
frustration occurring due to the interchain interactions.
\end{abstract}
]

Spin chains and ladders are of fundamental interest for solid state physics
due to their peculiar properties.\cite{dagotto} In the last years many 
spin-chain compounds were found, mostly cuprates~\cite{dagotto,kojima,mizumo} 
and vanadates.\cite{morre} There one can  distinguish between the 
corner-sharing compounds with a 180$^\circ$ (T)ransition metal--(L)igand--T 
bond (Sr$_2$CuO$_3$~\cite{kojima,rosner97}) and the edge-sharing compounds 
with a 90$^\circ$ T--L--T bond (Li$_2$CuO$_2$, CuGeO$_3$,\cite{mizumo} 
MgVO$_3$~\cite{morre}). In cuprates the relevant $d$ orbitals are directed 
towards the ligand ions, which results in a strong antiferromagnetic 
superexchange interaction for a 180$^\circ$ T--L--T bond and a weaker 
ferromagnetic coupling for a 90$^\circ$ T--L--T bond according to the
Goodenough-Kanamori-Anderson (GKA) rules.\cite{gka} Deviations from the GKA
rules are known for CuGeO$_3$ due to the presence of side 
groups.\cite{geertsma}

Recently, the compound MgVO$_3$ was proposed as a new candidate for a model
spin-chain system and magnetic susceptibility measurements were
presented.\cite{morre} The data suggest short range antiferromagnetic spin
correlations with the constant of the high-temperature Curie-Weiss law 
$\theta\approx-100$~K. The data were analyzed within a $1D$ spin-1/2 
Heisenberg model with the nearest neighbor $J_1$ and the next nearest neighbor
$J_2$ exchange couplings and a frustration $\alpha=J_2/J_1$ close to the 
critical value~\cite{okamoto} $\alpha_c=0.24$ was found. Here we present 
bandstructure calculations for this compound and a corresponding orbital and 
tight-binding (TB) analysis. 

The base-centered orthorhombic crystal structure of MgVO$_3$ was determined by
Bouloux {\em et al.}~\cite{bouloux} and is shown in Fig.~\ref{CS}. It consists
of edge-sharing VO$_2$ chains running along the $y$-direction which are 
coupled in $x$-direction by V--O--O--V bonds. For a convenient presentation of
the orbital analysis the coordinate system is rotated $90^\circ$ about the 
$x$-axis against the standard one~\cite{inttab} so that the space group reads 
as $Bm2_1b$ instead of $Cmc2_1$ given in Ref.~\onlinecite{bouloux}. 

\vbox{
\begin{figure}
\epsfxsize=\hsize\epsfbox{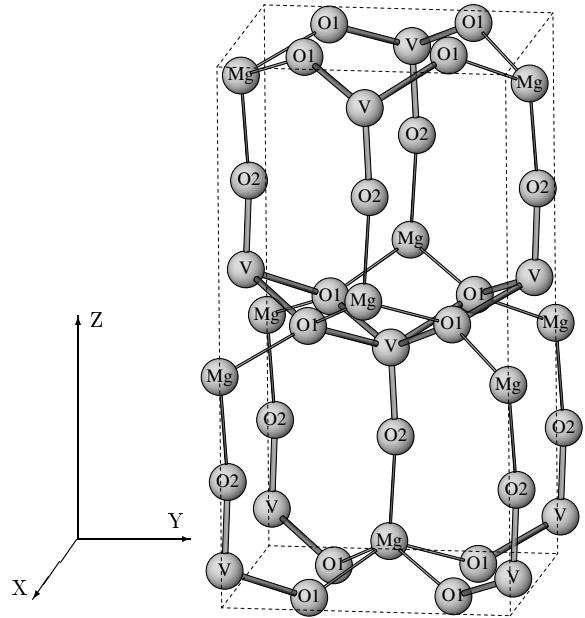}
\caption{Crystal structure of MgVO$_3$. The dashed lines depict the 
Bravais unit cell. The lattice parameters of the cell are 
$a=5.243$~\AA, $b=5.293$~\AA, $c=10.028$~\AA.}
\label{CS}
\end{figure}
}
\vskip-.75\baselineskip
One notes a slight tilting of the VO$_5$ pyramids out of the $z$-direction and
an asymmetric coordination of V and O1 ions. If one distorts the structure in 
a way depicted in Table~\ref{Wyckoff} both the tilting and the asymmetry are 
removed, a center of inversion appears, 
and the space group becomes $Bmmb$. All the following results were obtained 
using the model structure, because this simplification considerably reduces 
the computational efforts, and, as we have checked for the non-spin-polarized 
case, the deviation from the result obtained with the real crystal structure 
is negligible. 

\vbox{
\begin{table}
\caption{The Wyckoff positions of atoms in MgVO$_3$.}
\begin{tabular}{lcccccc}
&\multicolumn{3}{c}{Real structure}&\multicolumn{3}{c}{Model structure}\\
Atom&$x$&$y$&$z$&$x$&$y$&$z$\\
\tableline
\ Mg & 0 & 0 & \-0.4267 & 0 & 0 & 0.4267\\
\ V & 0 & 0.011 & \-0.0686 & 0 & 0 & 0.0686\\
\ O1 & 0.2383 & 0.264 & $-$0.005\0 & 0.2383 & 0.25 & 0\\
\ O2 & 0 & 0.025 & \-0.2330 & 0 & 0 & 0.2330\\
\end{tabular}
\label{Wyckoff}
\end{table}
}
The calculation of the bandstructure was performed using the full-potential 
nonorthogonal local-orbital minimum-basis bandstructure scheme (FPLO) within 
the local spin density approximation (LSDA).\cite{koepernik} The calculation 
was non-relativistic, the exchange and correlation potential was taken 
from Ref.~\onlinecite{perdew}. The set of valence orbitals was chosen to be 
Mg: $3s3p3d$, V: $3s3p4s4p3d$ and O: $2s2p3d$. The inclusion of the vanadium 
$3s3p$ states turned out to be unavoidable since the V--O2 distance of about 
$3.11\,a_0$ is small enough to yield a slight overlap of these states. The 
oxygen and magnesium $3d$ orbitals were taken to increase the basis 
completeness; though being not occupied they contribute to the overlap 
density. The extent of the basis orbitals, controlled by a confining potential
$(r/r_0)^4$, was optimized with respect to the total energy.

In our calculation the Bloch state $\ket{\k\nu}$ is composed of overlapping 
atomiclike orbitals $\ket{Lij}$ centered at the atomic sites $j$ in the unit 
cell $i$ with coordinates $\R_{ij}$:
\begin{equation}
\label{Bloch}
\ket{\k\nu}=\sum_{Lij} C^{\k\nu}_{Lij} e^{\i\k\R_{ij}}\ket{Lij},
\end{equation}
with the normalization condition $\langle\k\nu|\k\nu\rangle=1$. $L$ stands
for $\{nlm\sigma\}$ denoting the main, orbital, magnetic quantum numbers and
the projection of spin, respectively.

We have performed three kinds of computation: one non-spin-polarized and two 
spin-polarized with collinear and anticollinear spin polarization at the 
neighboring in-chain vanadium ions. In the last case we assumed ferromagnetic
order along the $x$- and $z$-directions. In both spin-polarized calculations 
we have found magnetic (ferro- and antiferro-) solutions with the total energy
being about 0.22~eV per formula unit lower than that of the paramagnetic 
solution of the non-spin-polarized calculation. The LSDA accuracy does not 
allow to determine which of the magnetic states is preferable. The magnetic 
moment $\langle n_\uparrow-n_\downarrow\rangle\,\mu_{\rm B}$ of the vanadium 
ion in both magnetic solutions is close to the saturated value 
$1\,\mu_{\rm B}$. The magnetic moments of the other ions are much smaller,
the only appreciable one occurs at apex oxygen (O2) having a value of about
$0.05\,\mu_{\rm B}$ and being directed opposite to the moment of the 
neighboring vanadium ion.

The results of all three calculations are presented in Fig.~\ref{BS}. The 
paramagnetic solution has metallic character with a half-filled conduction 
band at the Fermi level, whereas the band splits in two in the magnetic 
solutions and an insulator gap opens, being about 0.5~eV and 
\vbox{
\begin{figure}
\epsfxsize=\hsize\epsfbox{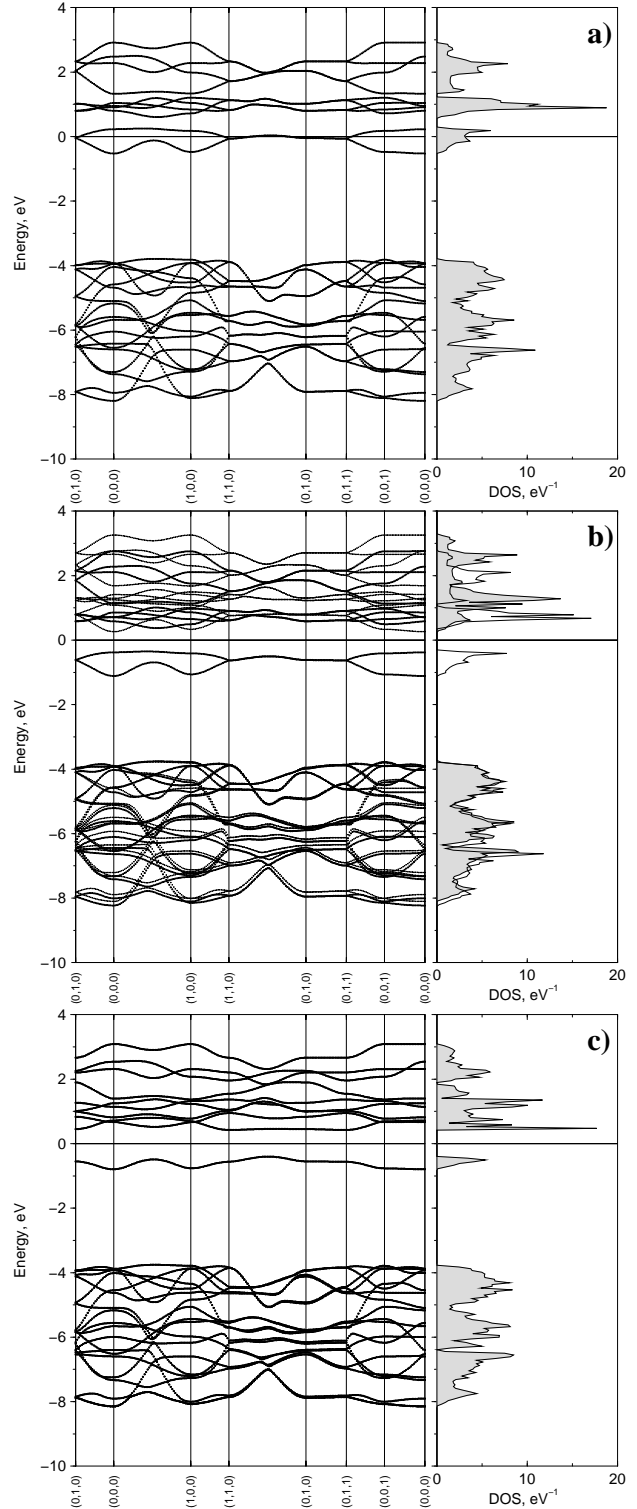}
\caption{Band structure and density of states of MgVO$_3$ for the paramagnetic
(a), ferromagnetic (b) and antiferromagnetic (c) solutions. The symmetry point
coordinates of the $B$-base centered orthorhombic Brillouin zone are given in 
units $\pi(2/a,1/b,2/c)$. The majority and minority (tinier dots and shaded 
DOS) spin parts are shown.}
\label{BS}
\end{figure}
}
0.8~eV in ferro- and antiferromagnetic cases, respectively.
However, it can be expected that the real gap is caused mainly by the electron 
correlation and is considerably larger than that one produced by the magnetic 
ordering. The correlation gap should persist in the paramagnetic case as well.

For a deeper understanding of the electronic structure we have performed an 
orbital analysis for the paramagnetic case. The weight $W^{\k\nu}_{Lj}$ of the 
orbital $\ket{Lj}$ in the Bloch state $\ket{\k\nu}$ (Eq.~\ref{Bloch}) was 
taken as
\begin{equation}
W^{\k\nu}_{Lj}=\sum_i |C^{\k\nu}_{Lij}|^2.
\end{equation}
The sum of all weights $\sum_{Lj} W^{\k\nu}_{Lj}$ is approximately unity, with 
some deviation due to the nonorthogonality of the basis orbitals. Actually the 
orbital weights were normalized with respect to the sum.

\begin{figure}
\epsfxsize=\hsize\epsfbox{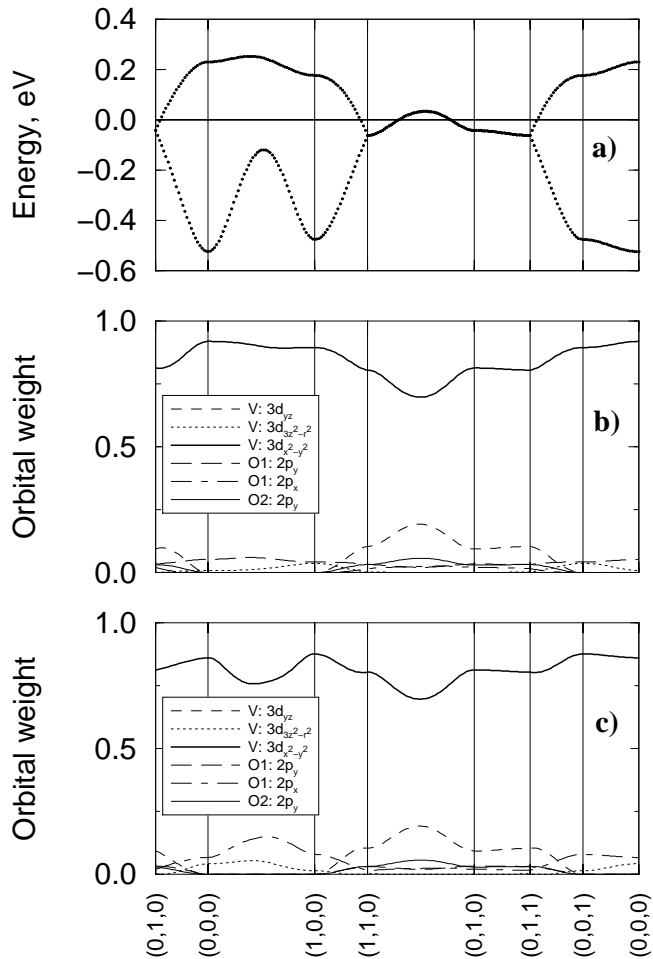}
\caption{Dispersion (a) and orbital weights of the upper (b) and lower (c)
parts of the band crossing the Fermi level in the paramagnetic solution. The
coordinates of $k$-points are given in the same units as in Fig.~\ref{BS}.}
\label{OW}
\end{figure}

According to the analysis, the valence bands consist of a lower oxygen 
$2p$-orbital complex which is separated by 3~eV from upper vanadium 
$3d$-orbitals (cf. Fig.~\ref{BS}). The $3d$-orbitals are split according to
the standard crystal field rules. At (0,1,0) their energy rises in the order: 
$d_{x^2-y^2}$, $d_{yz}$, $d_{zx}$, $d_{3z^2-r^2}$, $d_{xy}$. Only the lowest 
$3d_{x^2-y^2}$-orbital is partially occupied, being half-filled with two 
electrons per two vanadium atoms in the primitive unit cell. The corresponding
band (see Fig.~\ref{OW}) is very narrow having a width of only 0.8~eV. The 
weight of the V: $3d_{x^2-y^2}$ orbital in the band is larger than 70 percent,
but there are considerable contributions up to 20 percent from the O1: $2p_x$ 
and the V: $3d_{yz}$ orbitals, as well as smaller contributions from some 
other orbitals. 
Comparing the situation with that in cuprates having edge-sharing CuO$_2$ 
chains (Li$_2$CuO$_2$ or CuGeO$_3$) one observes an important distinction: 
in the cuprates the relevant Cu: $3d$-orbital is 
directed towards the oxygen ions, whereas in the present case it is directed 
towards the neighboring vanadium ions. It implies an indirect superexchange 
mechanism via the intermediate ligand in the cuprates but probably a dominant 
direct exchange process between neighboring vanadium ions and a much smaller 
superexchange hopping to the second in-chain neighbor in MgVO$_3$. Let us note
that the different coordination of the relevant $3d$-orbital in vanadates and 
cuprates with similar crystal structure has to be a rather common feature, 
because in vanadates the energetically lowest $3d$-orbital is half-filled in 
contrast to the highest one in cuprates. 

\begin{figure}
\epsfxsize=\hsize\epsfbox{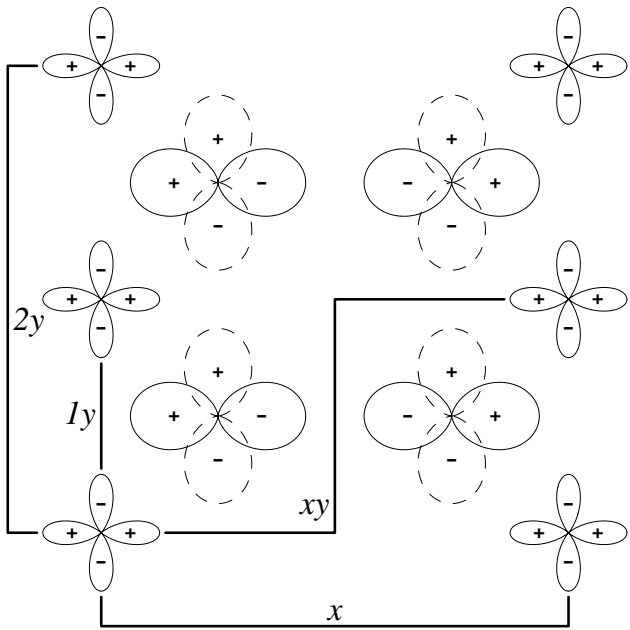}
\caption{Relevant V: $3d_{x^2-y^2}$ and O1: $2p_{x,y}$ orbitals coupling the 
spins in chain $y$-direction and in the orthogonal $x$-direction and the 
chosen in-plane transfer paths. The orbital phases correspond to the 
$\Gamma$-point.}
\label{OS}
\end{figure}

The coupling between the neighboring chains occurs mainly in the $xy$-plane 
via the intermediate O1: $2p_x$ orbitals. This is manifested by the strong 
dispersion along the $(k_x,0,0)$ direction (see Fig.~\ref{OW}) of the lower 
part of the band, which has a remarkable contribution from the orbital, 
whereas the upper part having a contribution only from the O1: $2p_y$ orbital
is almost dispersionless. Fig.~\ref{OS} showing the relevant orbitals at the 
$\Gamma$-point illustrates the interchain coupling via $\sigma$-bonds of 
neighboring O1: $2p_x$ orbitals.

To quantify the result a TB analysis has been performed for the relevant band
of the paramagnetic solution. The hopping processes to the nearest ($t_{1y}$)
and the next nearest ($t_{2y}$) in-chain neighbors, as well as two hopping 
processes ($t_{x},t_{xy}$) to the next chain and one ($t_{xyz}$) possible
coupling in the $z$-direction were taken into account:
\begin{eqnarray*}
E_0-E_{\bf k}=2t_{1y}\cos k_y\frac{b}{2}+2t_{2y}\cos{k_yb}+2t_x\cos{k_xa}\\
{}+4t_{xy}\cos{k_xa}\cos{k_y\frac{b}{2}}+
8t_{xyz}\cos{k_x\frac{a}{2}}\cos{k_y\frac{b}{2}}\cos{k_z\frac{c}{2}},
\end{eqnarray*}
yielding the parameters (in meV): $t_{1y}=125$, $t_{2y}=20 \ldots 26$, 
$t_{x}=50$, $t_{xy}=20$ and a tiny value of 3 for $t_{xyz}$. The data 
confirm that the ratio $t_{2y}/t_{1y}$, a measure for the frustration in the 
chain direction, is much smaller than the corresponding value for CuGeO$_3$
and Li$_2$CuO$_2$.\cite{weht,rosner99} It should be noted that the parameter 
$t_{1y}$ is much larger than the corresponding inter-ladder hopping in 
NaV$_2$O$_5$ given in Ref.~\onlinecite{smolinski}, probably due to a mutual 
compensation of different transfer paths in the ladder compound.

An estimate of the corresponding exchange integrals of the effective
Heisenberg Hamiltonian
$$
\hat{H}=\frac{1}{2}\sum_{ij} J_{j}\S_i\S_{i+j}
$$
can be performed in the one-band Hubbard description, which gives an 
antiferromagnetic exchange of roughly $J_j=4t_j^2/U$ with a value of $U$ 
still to be determined. Using the experimental value of 
$J_{1y}\approx 10$~meV given in Ref.~\onlinecite{morre} one gets $U=6.25$~eV 
which seems to be slightly overestimated. Most probably, additional 
ferromagnetic processes (possibly via the $3d_{yz}$ orbital or indirect 
processes via oxygen) have to be included to improve the estimate. Our TB 
analysis suggests a rather large value for $J_x$. It gives two-dimensional 
antiferromagnetic order with the frustration terms $J_{2y}$ and $J_{xy}$ 
(see Fig.~\ref{OS}). Though $t_{2y}$ and $t_{xy}$ are of nearly the same 
value, due to the larger coordination number, $J_{xy}$ dominates. In a simple 
mean field approach one can define an effective frustration exchange 
$J_{\rm eff}=2J_{xy}+J_{2y}$ yielding an effective frustration 
$J_{\rm eff}/J_{1y}$ of roughly $0.1$ which is a very preliminary analysis, 
however. 

\vskip-2\baselineskip

Thus, we have found that the VO$_2$ chains of MgVO$_3$ have indeed a spin-1/2 
structure. In contrast to the crystallographically similar edge-sharing 
cuprates the relevant $3d$ orbital is directed towards the nearest in-chain 
transition metal (V) ions. It suggests a direct antiferromagnetic exchange 
process between the neighboring in-chain spins. A similar process was proposed
earlier for the inter-ladder exchange coupling in MgV$_2$O$_5$.\cite{millet}
The importance of the direct vanadium $d$-$d$ transfer was also anticipated in
Ref.~\onlinecite{pickett} for CaV$_4$O$_9$, a compound with V-V distance 
slightly larger (3.00~\AA) than in MgVO$_3$ (2.96~\AA), where a value of 
80~meV for the transfer was reported. Besides the direct exchange we have 
found also some additional superexchange terms which give rise to the coupling
between the chains in the $x$-direction and to the frustration. The dominant 
frustration occurs between the neighboring chains in contrast to the naive 
picture of the frustration due to the next-nearest in-chain neighbor 
superexchange. It suggests that the experimental data require an analysis in 
the framework of a two- rather than a one-dimensional Heisenberg model.

\acknowledgments
We thank Christoph Geibel for useful discussions.

\end{document}